\documentclass[prb,twocolumn,showpacs,preprintnumbers,amsmath,amssymb]{revtex4}
\usepackage{graphicx}
\usepackage{dcolumn}
\usepackage{bm}

\begin{document}

\title{Sc substitution for Mg in MgB$_{2}$: effects on T$_{c}$ and Kohn anomaly}
\author{S. Agrestini, C. Metallo, M. Filippi, L. Simonelli, G. Campi, C. Sanipoli}
\affiliation{Dipartimento di Fisica, Universit\`{a} di Roma ``La
Sapienza", P. le Aldo Moro 2, 00185 Roma, Italy}
\author{E. Liarokapis}
\affiliation{Department of Applied Mathematics and Physics,
National Technical University of Athens, GR-157 80 Athens, Greece}
\author{S. De Negri, M. Giovannini, A. Saccone}
\affiliation{Dipartimento di Chimica e Chimica Industriale,
Universit\`{a} di Genova, Via Dodecaneso 31, 16146 Genova, Italy}
\author{A. Latini}
\affiliation{Dipartimento di Chimica, Universit\`{a} di Roma ``La
Sapienza", P. le Aldo Moro 2, 00185 Roma, Italy}
\author{A. Bianconi}
\affiliation{Unit\`{a} INFM and Dipartimento di Fisica,
Universit\`{a} di Roma ``La Sapienza", P. le Aldo Moro 2, 00185
Roma, Italy}\date{\today}
\begin{abstract}
    Here we report synthesis and characterization of
    Mg$_{1-x}$Sc$_{x}$B$_{2}$ (0.12$<$x$<$0.27) system, with critical temperature
     in the range of 30$>$T$_{c}$$>$6 K. We find that the Sc doping moves the chemical potential through
     the 2D/3D electronic topological transition (ETT) in the $\sigma$ band
     where the ``shape resonance" of interband pairing occurs. In the 3D regime beyond the ETT
     we observe a hardening of the E$_{2g}$ Raman mode with a significant
     line-width narrowing due to suppression of the Kohn anomaly over the range 0$<$q$<$2k$_{F}$.
\end{abstract}


\maketitle

Following the discovery of superconductivity in MgB$_{2}$ with a
T$_{c}$ of 40 K \cite{1a} large attention has been paid to
chemical substitutions \cite{2,3,4,5,6,7,8} aiming at enhancement
of T$_{c}$ and H$_{c2}$, and to manipulate the electronic
structure for the understanding of the high T$_{c}$
superconductivity. As a matter of fact, chemical substitution in
the MgB$_{2}$ is difficult and only Al replacing Mg \cite{4,5,6,7}
and C replacing B \cite{8} had been successful. Both of these
substitutions reduce the T$_{c}$ and induce a lattice compression.
The variation of T$_{c}$ with doping is mostly determined by the
tuning of the chemical potential through the electronic
topological transition (ETT) where the topology of the Fermi
surface of the $\sigma$ band changes from 2D to 3D \cite{6}. In
this two-gap superconductor \cite{9} the exchange-like non-diagonal
$\sigma$-$\pi$ interband pairing terms that enhance 
T$_{c}$ are expected to exhibit large variation near the ETT \cite{5,6}. 
In addition, Raman spectroscopy measurements on the Mg$_{1-x}$Al$_{x}$B$_{2}$ and
MgB$_{2-x}$C$_{x}$ systems \cite{10,11,12,13} have revealed a
line-width narrowing and energy hardening of the E$_{2g}$-mode
with the substitution beyond the ETT. This effect could be
interpreted in term of suppression of the Kohn anomaly
\cite{14,15} observed over an extended range 0$<$q$<$2k$_{F}$ in
the MgB$_{2}$ \cite{16,17} by a shift of the Fermi level. In fact
the Kohn anomaly \cite{18} is strong (weak) for a 2D (3D) Fermi
surface \cite{19,20}. Therefore it is expected to decrease for a
2D to 3D ETT \cite{6,20} of the Fermi surface. The proximity to an
ETT has been invoked to explain the anomalous pressure dependence
\cite{22} of this mode and it has been proposed to be the driving
mechanism for raising the critical temperature \cite{6,23,24}. The
aim of this work is to modify the band structure by chemical
substitutions to explore the role of electronic structure in the
MgB$_{2}$ on the electron-phonon coupling and on T$_{c}$ by tuning
the chemical potential through the shape resonance.

 We have synthesized the new superconducting ternary
system Mg$_{1-x}$Sc$_{x}$B$_{2}$ for 0.12$<$x$<$0.27, where the
chemical substitution induces minor lattice variations since the
ionic radius of Sc (1.62 \AA) is only a little larger than of Mg
(1.602 \AA). The substituted Sc ions donate 0.12$<$x$<$0.27
electrons per unit cell to the conduction $\sigma$ and $\pi$ bands
so the chemical potential is shifted towards the top of the
$\sigma$ band beyond the ETT \cite{6} where the $\sigma$ Fermi
surface changes from a 2D to a 3D topology, expected near x=0.12
\cite{9,18}. Sc substitution for Mg increases the disorder in the
Mg/Sc layers but has minor effects on the lattice structure of the
boron layer. Therefore the variation of T$_{c}$ and
electron-phonon coupling with the variation of the electron
structure can be well investigated.

 In these new samples we observe a remarkable line narrowing and
 frequency hardening of the E$_{2g}$ Raman mode, which gives a
 compelling experimental evidence for a drastic reduction of the
 Kohn anomaly and of the electron phonon coupling for x$>$0.12.
 It is remarkable to note that the critical temperature in the
 Mg$_{1-x}$Sc$_{x}$B$_{2}$ samples has been dropped only by a factor
 2-10 from MgB$_{2}$, much less than expected for a multi-band
 theory (with doping-independent interband pairing) considering the decrease of the electron-phonon
 coupling and of the density of states in the $\sigma$ band. This
 suggests importance of the resonant enhancement of the interband
 pairing term \cite{23} and the associated minimum of the non-diagonal
 coulomb pseudo-potential \cite{24} that drives the T$_{c}$ amplification
 at the ETT, called ``shape resonance" in a two-gap superconductor.

Mg$_{1-x}$Sc$_{x}$B$_{2}$ samples were synthesized by direct
reaction method of the elemental magnesium and scandium (powder,
99.9 mass \% nominal purity), boron (99.5 \% pure $<$60 mesh
powder). The starting materials were mixed in a stoichiometric
ratio and pressed into pellets of 8 mm in diameter. Each pellet
was enclosed in a tantalum crucible and sealed by arc welding
under argon atmosphere. The Ta crucibles were then heated in a
furnace Centorr M60 under high-pure Ar atmosphere for 14 hours in
the temperature range between 1280 and 950 $^{o}$C.

 The phase purity of the samples was checked by X-ray diffraction.
 The diffraction patterns of Mg$_{1-x}$Sc$_{x}$B$_{2}$ samples were
 measured in the Bragg-Brentano $\theta$-$\theta$ geometry by a vertical
X'Pert Pro MPD diffractometer using a Cu K$_{\alpha}$ radiation. The 
X-ray diffraction measurements of several samples were repeated at the 
beamline ID31 of the European Synchrotron Radiation Facility (ESRF), 
Grenoble. The samples were sealed in 1.0 mm diameter glass capillaries 
and the high-resolution diffraction profiles ($\lambda$=0.5 \AA) 
were collected at T=80 K using nine Ge(111) analyzer crystals.
The reflections were indexed to a MgB$_{2}$-like structure
according to the hexagonal AlB$_{2}$ structure type (P6/mmm space
group). No Sc, Sc$_{2}$O$_{3}$ or ScB$_{12}$ minority phases were
found, which indicates a successful Sc substitution for Mg. Figure
1a shows profiles of (002) and (110) diffraction peaks for
representative sample of the Mg$_{1-x}$Sc$_{x}$B$_{2}$ system. A
line broadening is observed in Sc-doped compounds as compared to
MgB$_{2}$ and ScB$_{2}$ indicating disorder or non-uniformities
due to Mg/Sc layers.

The samples were characterized for their superconducting
properties by the temperature dependence of complex conductivity
using the single-coil inductance method \cite{5,6}. The
temperature dependent radio-frequency complex conductivity for
representative Sc contents is shown in Fig. 1b where it can be
seen that the introduction of Sc in the Mg-planes induces a clear
shift of the superconducting transition to lower temperatures. The
superconducting transition for the Sc-doped samples shows a
broadening. This indicates that some disorder does exist in the
Sc-doped samples with possible effect on the superconductivity via
an increase of intraband scattering in the $\pi$ band.

The Raman spectra were measured in the back-scattering geometry,
using a T64000 Jobin-Yvon triple spectrometer with a
charge-coupled device camera. The explored Raman shift ranges
between 200 and 1100 cm$^{-1}$. The 488.0 nm Ar$^{+}$ laser line
was focused on 1-2 $\mu$m large crystallites and the power was
kept below 0.03 mW to avoid heating by the beam. Typical spectra
recorded for selected temperatures are shown in Fig. 1c. The
in-plane Boron vibration with E$_{2g}$ symmetry produces a single
narrow peak in Raman spectrum of ScB$_{2}$ as in AlB$_{2}$
\cite{10,12} also if at lower energy according with its larger
lattice parameters. The Raman line is softened going from
ScB$_{2}$ to Mg$_{1-x}$Sc$_{x}$B$_{2}$ and finally becomes very
soft and very broad in MgB$_{2}$.

The diffraction data were analyzed by Rietveld refinement using
GSAS program. The behaviour of the lattice parameters $a$ and $c$
as a function of Sc content is reported in Fig. 2. A miscibility
gap occurs in the range from 2$\pm$1\% to 12$\pm$1\% Sc
substitution where the samples show a macroscopic phase separation
between low-doped and high-doped samples. The $a$-axis increases
gradually with increasing Sc-content, while the $c$-axis is nearly
constant. The unit cell volume shows a small expansion in
agreement with the similar ionic radius between Sc$^{3+}$ and
Mg$^{2+}$. For comparison in the same figure we report the
evolution of the lattice parameters in the
Mg$_{1-x}$Al$_{x}$B$_{2}$ system that shows much larger lattice
variations. Here ``x" is the nominal Sc-content and we can
conclude that the introduced Sc is successfully substituted for
Mg, and the actual Sc-content is not much different from the
nominal value. From the variation of lattice parameters with Sc
content, we can conclude that the solubility range of Scandium in
MgB$_{2}$ is between 12\% and 27\%.

The variation of the superconducting critical temperature T$_{c}$
as a function of Sc doping is reported in the panel (a) of Fig. 3.
The transition temperature T$_{c}$ was determined from the peak in
derivative of the complex conductivity. The T$_{c}$ decreases
continuously with increasing Sc substitution from 30K at 12\% to
6K at 27\%. It is interesting to note that the variation of
T$_{c}$(x) is sharper for 0.12$<$x$<$0.15.

The panel (b) of Fig. 3 shows the variation of the frequency of
the E$_{2g}$ Raman line from MgB$_{2}$ to
Mg$_{1-x}$Sc$_{x}$B$_{2}$. The error bars indicate the E$_{2g}$
peak half-width. In the Sc-substituted samples in the range
12\%-27\% the phonon peak is shifted toward higher frequency, it
is much narrower in comparison with MgB$_{2}$ and it is
approaching the frequency of that for the ScB$_{2}$. The frequency
hardening and the line-width narrowing of the Raman E$_{2g}$ mode
indicates a clear decrease of the electron-phonon coupling going
from MgB$_{2}$ to the Sc-doped samples.

Finally in Fig. 4 we report the energy of the Raman E$_{2g}$ mode
as a function of the $a$-axis (that is the relevant parameter for
the in plane high frequency longitudinal optical mode E$_{2g}$) of
the non-superconducting diborides ScB$_{2}$ and AlB$_{2}$, and of
the superconducting MgB$_{2}$, Mg$_{1-x}$Sc$_{x}$B$_{2}$ and
Mg$_{0.5}$Al$_{0.5}$B$_{2}$ \cite{13} systems.

Let us consider first the AlB$_{2}$ and ScB$_{2}$ samples with a
filled $\sigma$ band. The energy of the E$_{2g}$ mode as a
function of $a$-axis follows the law
$\Omega(a)=\omega_{Al}(a/a_{Al})^{-3\gamma_{0a}}$ (dashed line in
panel a) where $\gamma_{0a}$ is the Gruneisen parameter
$\gamma_{0a}=- \partial\ln\nu/3\partial\ln a=1.4\pm$0.1 that is
the expected behaviour due to lattice expansion for a metallic
covalent material.

In the case where the Fermi level is tuned below the top of the
$\sigma$ band (e.g. MgB$_{2}$, Mg$_{1-x}$Sc$_{x}$B$_{2}$ and
Mg$_{0.5}$Al$_{0.5}$B$_{2}$) a phonon decay channel opens up
unlike others with filled $\sigma$ band (e.g. AlB$_{2}$ and
ScB$_{2}$). In fact the phonons can now decay into electron-hole
excitations in the $\sigma$ band inducing a phonon energy
softening and line broadening (Kohn anomaly). The E$_{2g}$ phonon
softening can be obtained by the energy difference between the
experimental E$_{2g}$ Raman energy $\omega_{E_{2g}}$ and the
expected phonon energy $\Omega$$_{E_{2g}}$(a) for a material with
a filled $\sigma$ band and with the appropriate lattice parameter.
We deduce the large phonon softening of 37 meV for MgB$_{2}$ and
about 17 meV for Mg$_{1-x}$Sc$_{x}$B$_{2}$ system that is close to
the case of Mg$_{0.5}$Al$_{0.5}$B$_{2}$. The Raman softening in
MgB$_{2}$ is due to large electron-phonon interaction with the
electron-hole excitations in the $\sigma$ band that drives the
system close to a lattice instability and the breakdown of Migdal
approximation \cite{9,18}. The decrease of the electron-phonon
coupling near q=0 going from MgB$_{2}$ to
Mg$_{1-x}$Sc$_{x}$B$_{2}$ is given by the variation of the ratio
$(\Omega_{E_{2g}}(q=0)/\omega_{E_{2g}}(q=0))^{2}$ shown in panel
(b) going from 2.2 in MgB$_{2}$ to 1.4 in
Mg$_{0.8}$Sc$_{0.2}$B$_{2}$. This drastic decrease is related to a
smaller E$_{2g}$ Kohn anomaly in Mg$_{1-x}$Sc$_{x}$B$_{2}$. This
result is confirmed by the variation of the ratio between the
line-width and the phonon energy 2$\gamma/\omega_{E_{2g}}$ shown
in panel c of Fig. 4 that decreases going from MgB$_{2}$ to
Mg$_{0.8}$Sc$_{0.2}$B$_{2}$. These results can be understood if
the Sc substitution has driven the chemical potential through the
ETT where the $\sigma$ Fermi surface has a 3D topology \cite{6}
with a reduced Kohn anomaly \cite{14,15}.

In conclusion we have reported the successful substitution of Sc
for Mg with 0.12$<$x$<$0.27 obtaining a very small lattice
expansion. The system shows a miscibility gap in the range
0.02$<$x$<$0.12, which supports the fact that the system is in the
proximity of a lattice instability expected at the 2.5 Lifshitz
phase transition \cite{21} in agreement with pressure effect
measurements \cite{22}. The effect of the Sc doping is to shift
the chemical potential toward the ETT and the top of the $\sigma$
band. However the rigid band model is not appropriate since we
expect that the Sc substitution increases the sp$^{2}$(B)-d(Sc)
hybridization and the dispersion of the $\sigma$ band.

In the Mg$_{1-x}$Sc$_{x}$B$_{2}$ samples the E$_{2g}$ Raman mode
shows a large hardening and narrowing in comparison with MgB$_{2}$
while the lattice expands, that indicates a drastic reduction of
the Kohn anomaly in the E$_{2g}$ longitudinal optical mode. This
effect has been associated with the tuning of the Fermi level
beyond the critical energy for the ETT where the two dimensional
topology of the $\sigma$ Fermi surface changes to a three
dimensional topology. The critical temperature in the
Mg$_{1-x}$Sc$_{x}$B$_{2}$ samples has been found in the range 30-6
K as in the case of Al substituted samples in the regime where the
Fermi surface of the $\sigma$ band has a 3D topology beyond the
2D/3D ETT \cite{6}. The high values of T$_{c}$ in this multi-band 
superconductor are associated with the key role of the
non-diagonal interchannel pairing term and coulomb
pseudo-potential term as in the case of Al substitution
\cite{23,24}.
These new data support the key role of the
electronic structure controlling the resonant enhachement of the
exchange like interband coupling term in a two-gap superconductor.
This occurs when the chemical potential is tuned in an energy
window around the 2D/3D electronic topological transition in one
of the two Fermi surface portions giving the shape resonance \cite{25}
that pushes up the critical temperature.

We thank the ESRF for provision of synchrotron radiation facilities 
and we would like to thank Prof. A. Fitch and Dr. I. Margiolaki 
for assistance in using beamline ID31. We 
thank R. de Coss for useful discussions. This work was
supported by MIUR in the frame of the project Cofin 2003
``Synthesis and properties of new borides", by ``Istituto
Nazionale Fisica della Materia" (INFM), and by ``Consiglio
Nazionale delle Ricerche" (CNR) in the frame of the project 5\%
``Applicazioni della superconduttivit\`{a} ad alte T$_{c}$" law
95/95.


\begin{figure*}
\input{epsf}
\epsfxsize 6.5cm \centerline{\epsfbox{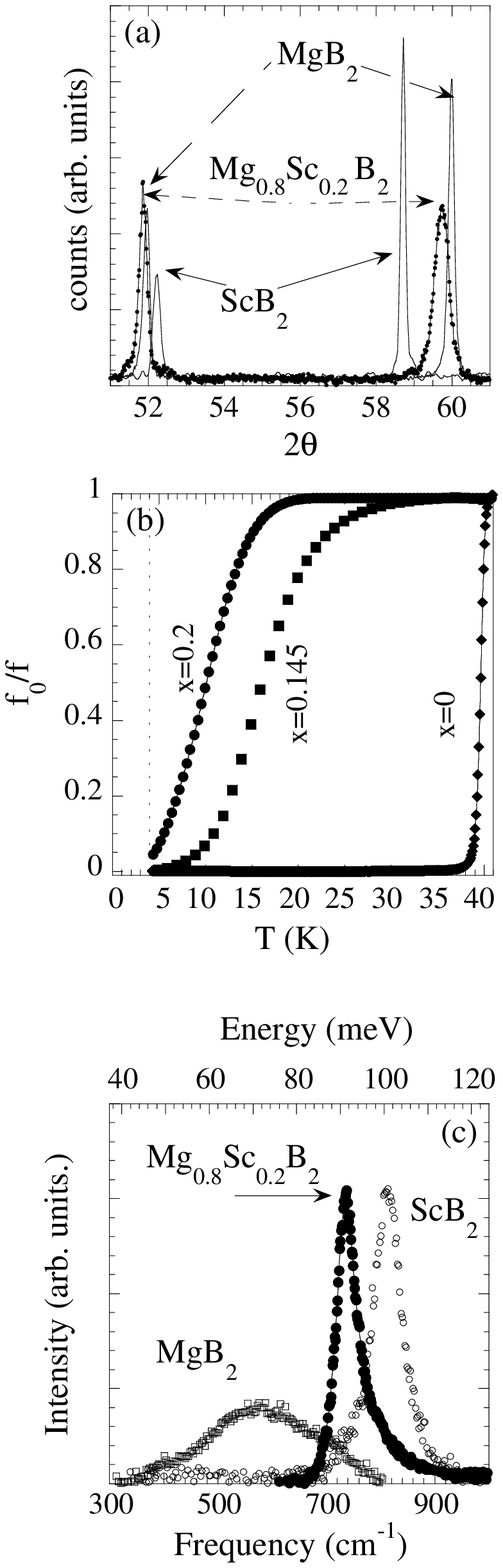}}
\caption{\label{fig:epsart} The
X-ray diffraction patterns in the range of the (001) and (110)
reflections (panel a), the superconducting transition measured by
complex resistivity (panel b), and the Raman spectra (panel c) are
shown for representative Mg$_{1-x}$Sc$_{x}$B$_{2}$ samples
compared with MgB$_{2}$ and ScB$_{2}$.}
\end{figure*}

\begin{figure*}
\input{epsf}
\epsfxsize 6.5cm \centerline{\epsfbox{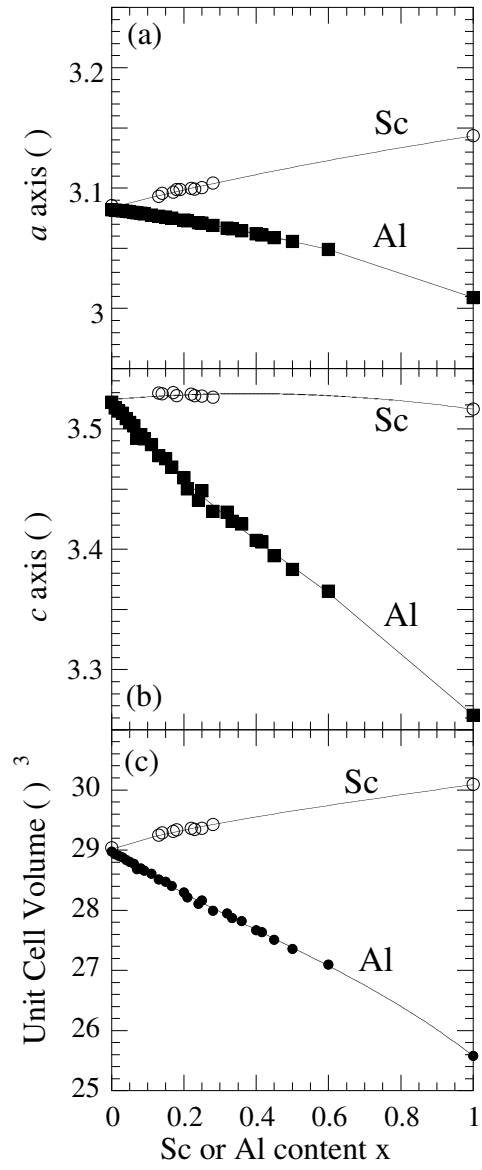}}
\caption{\label{fig:epsart} Changes of lattice parameters $a$
(panel a) and $c$ (panel b) and of unit cell volume (panel c) as a
function of Sc content x in Mg$_{1-x}$Sc$_{x}$B$_{2}$.}
\end{figure*}

\begin{figure*}
\input{epsf}
\epsfxsize 6.5cm \centerline{\epsfbox{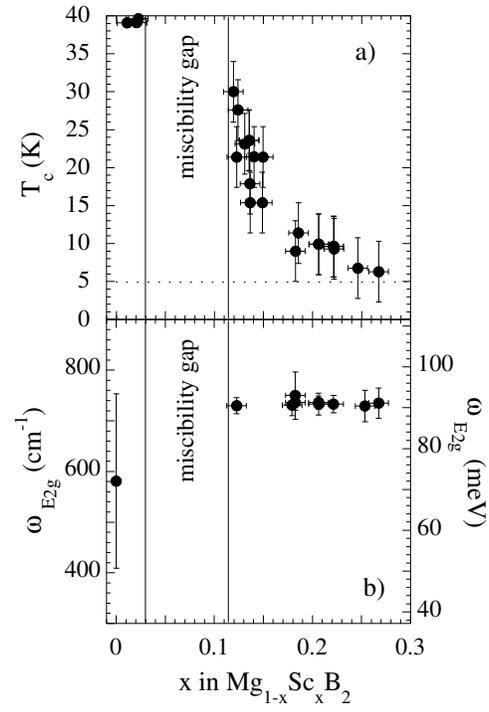}}
\caption{\label{fig:epsart} Evolution of the superconducting
critical temperature T$_{c}$ (panel a), the frequency
$\omega_{E_{2g}}$ (panel b) of the E$_{2g}$ Raman line as a
function of x in Mg$_{1-x}$Sc$_{x}$B$_{2}$. The superconductive
transition width and the E$_{2g}$ peak half-width are indicated as
error bars respectively in panel a and panel b.}
\end{figure*}

\begin{figure*}
\input{epsf}
\epsfxsize 6.5cm \centerline{\epsfbox{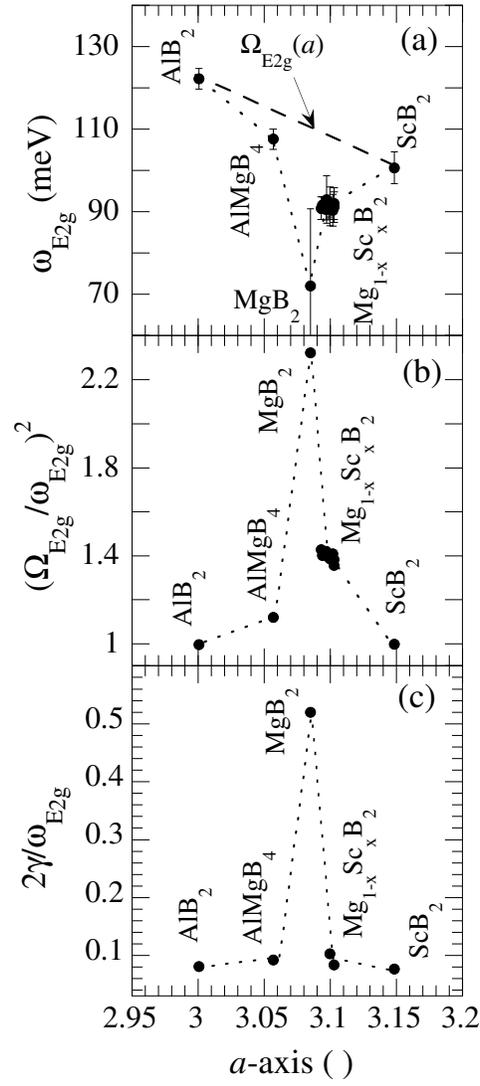}}
\caption{\label{fig:epsart} Variation of the frequency of the
E$_{2g}$ Raman line as function of the lattice parameter $a$ for
different diborides (panel a). The softening and broadening of the
E$_{2g}$ mode due to the Kohn anomaly is given by the energy ratio
($\Omega_{E_{2g}}(a)/\omega_{E_{2g}}(a))^{2}$ (panel b) and the
ratio 2$\gamma/\omega_{E_{2g}}$ (panel c) where 2$\gamma$ is the
Raman line-width.}
\end{figure*}

\end{document}